\documentclass[twocolumn ,aps,prd,a4paper,superscriptaddress,tightenlines,noeprint,11pt,floatfix]{revtex4-2}
\RequirePackage{float}
\usepackage[normalem]{ulem}
\usepackage[utf8]{inputenc}
\usepackage[USenglish]{babel}
\usepackage[T1]{fontenc}
\usepackage{graphicx,epsf}
\usepackage[toc,page,header]{appendix}
\usepackage{minitoc}
\usepackage{color}
\usepackage[bookmarksnumbered,hypertexnames=false,colorlinks,colorlinks=true,linkcolor=blue,urlcolor=black,citecolor=blue,anchorcolor=green]{hyperref}
\usepackage{bbm,braket,microtype,mathrsfs,amsmath,amssymb,color,amsthm,mathtools,fullpage,enumitem,ae,aecompl,bm,thm-restate}
\usepackage{tikz}
\usepackage{lipsum}
\usepackage{nameref}
\usepackage{color}
\usepackage[capitalize]{cleveref}
\usepackage{silence}
\usepackage{mathtools}

\allowdisplaybreaks
\newcommand{\be}{\begin{equation}}
\newcommand{\ee}{\end{equation}}
\newcommand{\ba}{\begin{eqnarray}}

\newcommand{\half}{\frac{1}{2}}
\newcommand{\ea}{\end{eqnarray}}
\DeclareMathOperator{\tr}{tr}

\newcommand\stab{{\operatorname{STAB}}}
\newcommand{\ignore}[1]{}

\newcommand{\sswap}{\operatorname{T}}

\newcommand{\ot}{\otimes}
\newcommand{\ZZ}{\mathcal{Z}_2}
\newcommand{\pauli}[1]{\mathbb{P}_{#1}}

\newcommand{\ketbra}[1]{\ket{#1}\hspace{-0.1 cm} \bra{#1}}

\newcommand{\absval}[1]{\left| {#1} \right|}

\newcommand{\Id}{\mathbbm{1}}

\newcommand{\hilbert}[1]{\mathcal{H}_{{#1}}}
\newcommand{\hi}{\mathcal{H}}

\newcommand{\poly}{\operatorname{poly}}

\newcommand{\sym}{\text{sym}}

\newcommand{\Y}{\operatorname{Y}}

\newcommand{\X}{\operatorname{X}}
\newcommand{\Z}{\operatorname{Z}}

\newcommand{\exv}{\mathbb{E}}
\newcommand{\lin}{\operatorname{lin}}
\newcommand\pdfmath[1]{\texorpdfstring{$#1$}{#1}}

\def\norm#1{\Vert #1\Vert}
\def\CC{{\rm\kern.24em \vrule width.04em height1.46ex depth-.07ex
   \kern-.29em C}}
\def\P{{\rm I\kern-.25em P}}
\def\RR{{\rm
        \vrule width.04em height1.58ex depth-.0ex
        \kern-.04em R}}

\def\bbbc{{\mathchoice {\setbox0=\hbox{$\displaystyle\rm C$}\hbox{\hbox
to0pt{\kern0.4\wd0\vrule height0.9\ht0\hss}\box0}}
{\setbox0=\hbox{$\textstyle\rm C$}\hbox{\hbox
to0pt{\kern0.4\wd0\vrule height0.9\ht0\hss}\box0}}
{\setbox0=\hbox{$\scriptstyle\rm C$}\hbox{\hbox
to0pt{\kern0.4\wd0\vrule height0.9\ht0\hss}\box0}}
{\setbox0=\hbox{$\scriptscriptstyle\rm C$}\hbox{\hbox
to0pt{\kern0.4\wd0\vrule height0.9\ht0\hss}\box0}}}}
\def\bbbz{{\mathchoice {\hbox{$\sf\textstyle Z\kern-0.4em Z$}}
{\hbox{$\sf\textstyle Z\kern-0.4em Z$}}
{\hbox{$\sf\scriptstyle Z\kern-0.3em Z$}}
{\hbox{$\sf\scriptscriptstyle Z\kern-0.2em Z$}}}}

\newlength{\fighskip} \fighskip=2pt
\newlength{\figvskip} \figvskip=1pt

\makeatletter
\def\namedlabel#1#2{\begingroup
   \def\@currentlabel{#2}%
   \label{#1}\endgroup
}
\makeatother

\usepackage{hyperref}

\newtheorem{lemma}{Lemma}

\begin{document}
\setcounter{secnumdepth}{3}
\title{Stabilizer entropy of quantum tetrahedra}
\author{Simone Cepollaro}
\email{simone.cepollaro-ssm@unina.it}
\affiliation{Scuola Superiore Meridionale, , Largo S. Marcellino 10, 80138 Napoli, Italy}
\affiliation{INFN, Sezione di Napoli, Italy}
\author{Goffredo Chirco}\email{goffredo.chirco@unina.it}
\affiliation{INFN, Sezione di Napoli, Italy}
\affiliation{Dipartimento di Fisica `Ettore Pancini', Universit\`a degli Studi di Napoli Federico II,
Via Cintia 80126,  Napoli, Italy}

\author{Gianluca Cuffaro}\email{gianluca.cuffaro001@umb.edu}
\affiliation{Physics Department,  University of Massachusetts Boston,  02125, USA}
\author{Gianluca Esposito} \email{g.esposito@ssmeridionale.it}
\affiliation{Scuola Superiore Meridionale, , Largo S. Marcellino 10, 80138 Napoli, Italy}
\author{Alioscia Hamma}\email{alioscia.hamma@unina.it}
\affiliation{Scuola Superiore Meridionale, , Largo S. Marcellino 10, 80138 Napoli, Italy}
\affiliation{INFN, Sezione di Napoli, Italy}
\affiliation{Dipartimento di Fisica `Ettore Pancini', Universit\`a degli Studi di Napoli Federico II,
Via Cintia 80126,  Napoli, Italy}

\begin{abstract}
How complex is the structure of quantum geometry? In several approaches, the spacetime atoms are obtained by the $SU(2)$ intertwiner called quantum tetrahedron. The complexity of this construction has a concrete consequence in recent efforts to simulate such models and toward experimental demonstrations of quantum gravity effects. There are, therefore, both a computational and an experimental complexity inherent to this class of models. In this paper, we study this complexity under the lens of {\em Stabilizer Entropy} (SE). We calculate the SE of the gauge-invariant basis states and its average in the $SU(2)-$gauge invariant subspace.  We find that the states of definite volume are singled out by the (near) maximal SE and give precise bounds to the verification protocols for experimental demonstrations on available quantum computers. 
\end{abstract}
\maketitle

\section{Introduction}
Understanding whether gravity admits a quantum formulation is one of the most intriguing challenges of modern physics. In the last decade, quantum information theory has provided new conceptual and mathematical tools to investigate the structure of spacetime at the quantum scale. Entanglement entropy has been used to probe the holographic architecture of spacetime, supporting the idea of entanglement as an essential  \emph{resource} to the emergence of classical spacetime geometry~\cite{RT, VanRaamsdonk:2010pw, MaldSuss13, Bianchi_2014, PhysRevD.95.024031}. Recently, also fostered by new perspectives in quantum gravity phenomenology, the use of quantum information tools to design and investigate \emph{experimental evidence} for quantum features of the gravitational field has attracted much attention~\cite{westphal2021measurement,Overstreet_2023, Christodoulou_2023}. 

Within the limits of current experimental technology, the first widely available quantum computers today allow to \emph{simulate} quantum gravity states, providing suggestions, predictions and setup ideas for future experiments~\cite{Czelusta_2021,vanderMeer:2022jec, polino2022photonic}. 
 In this scenario, we expect {\it non-stabilizer resources} to play a {double} key role. Gently speaking, non-stabilizerness is a core property of quantum states describing the complexity of the expression of their density operator in a specific operator basis (in the case of qubit systems, the Pauli operator basis), and its interplay with entanglement is known to be the essential ingredient needed to unlock quantum advantage~\cite{Campbell_2017, tirrito2023quantifying}.
Recently, this resource has been shown to be given an entropic meaning, as Stabilizer Entropy (SE), making it both computable~\cite{leone2022StabilizerRenyiEntropy} and measurable~\cite{Oliviero:2022bqm}. 
SE directly affects the cost (in terms of classical resources) of simulating a quantum state or process:  a $n$-qubit state or circuit using a number $t$ of non-stabilizer resources can be simulated with a classical computer at a computational cost that scales as $\exp(t)\poly(n)$\cite{aaronson2004ImprovedSimulationStabilizer}.  In particular, it provides bounds on the fidelity reachable in experimental realizations of quantum states \cite{nonstabfidest}. {Moreover, SE is involved in the onset of universal, complex patterns of entanglement~\cite{piemontese2022EntanglementComplexityRokhsarKivelsonsign,true2022TransitionsEntanglementComplexity}, quantum chaos~\cite{leone2021IsospectralTwirlingQuantum,oliviero2021RandomMatrixTheory,leone2021QuantumChaosQuantum,oliviero2021TransitionsEntanglementComplexity}, complexity in the wave function of quantum many-body systems~\cite{oliviero2022MagicstateResourceTheory,haug2023QuantifyingNonstabilizernessMatrix}, and decoding algorithms from black hole's Hawking radiation~\cite{leone2022RetrievingInformationBlack,oliviero2022BlackHoleComplexity,leone2022LearningEfficientDecoders, hayden2007BlackHolesMirrors}.}
States and processes with high non-stabilizer resources are generically exponentially harder to simulate on a classical computer, and they are harder to certificate in experimental protocols. {An} analysis of such resources is therefore vital in order to assess the simulability of quantum gravity states. {At the same time}, we expect such property to provide a new tool, in addition to entanglement, to investigate the emergence of classical spacetime in quantum gravity.

In this work, we explore the novel direction of looking at the non-stabilizerness of quantum gravity states in the setting of non-perturbative theories. We consider a description of quantum geometry given in terms of \emph{spin network} states, a general tool shared by lattice gauge theory~\cite{Donnelly_2014} and several background-independent approaches to quantum gravity (like loop quantum gravity~\cite{Thiemann_2007, rovelli_vidotto_2014}, state sum models~\cite{ooguri92, baez1999quantum}, and group field theories~\cite{oriti2007group,oriti2009group,oriti2015group}), where they provide a gauge-invariant basis for the field. 
A spin network is represented by a graph $\Gamma$, with edges and nodes colored respectively by $SU(2)$ spin halves and intertwiner operators~\cite{rovelli_vidotto_2014}.

Each node of the network graph is dual to a quantum polyhedron geometry, with a number of faces equal to the valence of the node. 
We focus on a single $4-$valent node, that is an $SU(2)$-gauge invariant state corresponding to a quantum tetrahedron~\cite{Ponzano1969,Regge1961, BARBIERI1998714, Bianchi_2011}.
 
We study the non-stabilizerness of quantum tetrahedron states using Stabilizer Entropy (SE) \cite{leone2022StabilizerRenyiEntropy}. The first  result of the work is that the states that diagonalize the oriented volume operator are the ones with highest value of SE: this result provides a new lower bound for the number of preparations needed in future experimental setups of quantum gravity states. To our knowledge, at present, this bound, obtained from the calculated intertwiner states non-stabilizerness has not been fully reached yet (see e.g. the data from the experiments realized in \cite{Czelusta_2021}). 
As a second result, we show that the projection into the gauge invariant Hilbert space associated to the process of constructing the quantum tetrahedron out of a collection of four qubits inherently requires non-stabilizer resources. Such resources become an intrinsic feature of the quantum geometry state, reflecting in the computational complexity of a simulation of such processes, in a way that is ultimately dependent on the structure of the gauge invariant space itself.

The paper is organized as follows. 
\cref{stabform} provides the basic notions of the stabilizer formalism necessary for our analysis. We recall the definition of Pauli operators, the Clifford group and construct the set of pure stabilizer states, highlighting the necessity to go beyond this set of states in order to unlock quantum advantage. Then, we introduce the definition of Stabilizer Entropy as an entropic measure of nonstabilizerness of a pure quantum state, as well as its properties. 
In \cref{sec:quantumtet} we introduce the setting of quantum gravity. We realize a quantum tetrahedron via projection into the SU$(2)$ gauge invariant (intertwiner) subspace of a spin network Hilbert space. We show the most general intertwiner state for any SU$(2)$ spin$-j$ irrep and then focus on the case of $j=1/2$. In \cref{sec:SEqt} and \cref{sec:simulations} 
we compute the non-stabilizerness of the logical basis and of the volume eigenstates basis elements. Then, these numerical results provide an estimate of the upper bound of the fidelity of the experiment in \cite{Czelusta_2021}; we compare our estimations with the experimental fidelity obtained. In \cref{sec:gap} we extend the analysis to non-stabilizerness of subspaces, in order to investigate the cost of projection in terms of non-stabilizer resources. To this end, we introduce the average SE gap onto a subspace, and we show that this quantity is directly dependent from the internal structure of said subspace in the form of its projector. Finally, we apply our obtained results to the intertwiner subspace and conclude that imposing the $SU(2)-$gauge invariance has an intrinsic cost in terms of non-stabilizer resources.

\section{Stabilizer formalism and Stabilizer Entropy}\label{stabform}
In this section, we review the stabilizer formalism and its role in the quantum computation framework.

    Let $\hi\simeq\mathbb{C}^{2\otimes n}$ a $n$-qubit system and $\pauli{n}$ be the Pauli group acting on $\hi$. Define the \textit{Clifford group} $\mathcal{C}(n)\subset \mathcal{U}(n)$ as the normalizer of the Pauli group, namely, $\mathcal{C}(n):=\{ C\in \mathcal{U}(n)\,,\, \text{s.t. }\,\forall P\in \pauli{n}\,,CPC^{\dag}=P'\in \pauli{n}\}$ \cite{gottesman1998heisenberg}. Hence, given a \textit{computational basis} $\{\ket{i}\}$ of $\hi$ as the common eigenbasis of the operators belonging to $\ZZ:=\{\Id,Z\}^{\otimes n}$, one can define the set of pure \textit{stabilizer states} of $\hi$ as the full Clifford orbit of $\{\ket{i}\}$\cite{veitch2014ResourceTheoryStabilizer}, namely
\begin{equation}
        \stab=\{C\ket{i}\,,C\in \mathcal{C}(n)\}\,.
    \end{equation}
    Stabilizer states share some properties with regard to the computational complexity of simulating quantum processes using classical resources; these properties are summarized by the Gottesman-Knill theorem, which states that any quantum process that can be represented with initial stabilizer states upon which one performs (i) Clifford unitaries, (ii) measurements of Pauli operators, (iii) Clifford operations conditioned on classical randomness, can be perfectly simulated by a classical computer in polynomial time \cite{gottesman1998heisenberg}. This means that stabilizer states and Clifford operators are not actually ``quantum" from a computational perspective, since they do not provide any advantage over classical computers. Since the set of stabilizer states is by definition closed under Clifford operations, a certain amount of resources beyond the Clifford group is needed to prepare a generic state in the Hilbert space: this quantity is referred to as \textit{non-stabilizerness} of this state, which has been proven to be a useful resource for universal quantum computation \cite{bravyi2005UniversalQuantumComputation} and for which several measures have been proposed \cite{howard2017ApplicationResourceTheory, veitch2014ResourceTheoryStabilizer}. For our analysis, we are going to use two entropic non-stabilizerness measures called \textit{2-Stabilizer R\'enyi Entropy (SE)} \cite{leone2022StabilizerRenyiEntropy} and its linear counterpart. They are defined starting from the probability distribution $\Xi_P(\ket{\psi}):=d^{-1}\tr^2(P\ketbra{\psi})$, with $P\in\pauli{n}$ and $d=\operatorname{dim}(\hi)=2^n$, associated to the tomography of the quantum state $\psi$. Then, the $2$-Stabilizer R\'enyi Entropy for pure states is defined as 
\begin{equation}
        \begin{split}
            M_2(\ket{\psi})&:=-\log{d\norm{\Xi_P(\ket{\psi})}^2 _2}\\
            &=-\log{d^{-1}\sum_{P\in\pauli{n}}\tr^4(P\ketbra{\psi})}\,,
        \end{split}
    \end{equation}
    whereas the linear SE is defined as
    \begin{equation}
        \begin{split}
            M_{\lin}(\ket{\psi})&:=1-d\norm{\Xi_P(\ket{\psi})}^2 _2\\
            &=1-d^{-1}\sum_{P\in\pauli{n}}\tr^4(P\ketbra{\psi})\,.
        \end{split}
    \end{equation}
Both $M_2$ and $M_{\lin}$ are: (i) faithful, i.e. $M(\psi)=0 \Leftrightarrow \psi \in \stab$, otherwise $M(\psi)>0$; (ii) invariant under Clifford operators, namely $M(C\psi C^\dag)=M(\psi)$, whereas $M_2$ is also additive under tensor product of quantum states \cite{leone2022StabilizerRenyiEntropy}. Both measures can also be written in a more compact form:
\begin{equation}\label{sredef}
    \begin{split}
        M_2(\ket{\psi})&=-\log{d \tr Q\psi^{\ot 4}}\,,\\
        M_{\lin}(\ket{\psi})&=1- d\tr Q \psi^{\ot 4}\,,
    \end{split}
\end{equation}
with $Q:=d^{-2}\sum_{P\in \pauli{n}} P^{\ot 4}$. 
This construction can be generalized in a straightforward way to the Pauli group (namely, the Heisenberg-Weyl group) for qudits, that is, $l-$level systems \cite{Wang2023}, and the SE are defined exactly as in Eq.\eqref{sredef}, see \cref{appqudit} for details.

Stabilizer entropies quantify the computational complexity of qubit states by the entropy of the distribution over the Pauli basis: states with high values of $M_2$ or $M_{\lin}$ require an exponential amount of classical resources to be simulated, and hence are those which may exhibit quantum advantage. 

Moreover, a  result shown in \cite{nonstabfidest} establishes a bound between SE and minimum number of copies needed to achieve a certain fidelity in certification protocols: 
therefore, from a computational perspective, the knowledge of SE of quantum gravity states is an essential tool to optimize time and resources involved in the preparation of the states on a quantum computer, once a desired value of fidelity has been established.

\section{Quantum tetrahedron}\label{sec:quantumtet}

Spin networks are symmetric tensor network states defined by a graphs $\Gamma$, labeled by SU$(2)$ irreducible representations and intertwining operators. In loop quantum gravity, such states encode the quantum description of the 3D space manifold into purely combinatorial and algebraic variables~\cite{PhysRevD.52.5743, PhysRevD.82.084040}. More generally, spin networks can be defined as abstract quantum many-body-like collections of \emph{fundamental} quanta of space, connected by maximally entangled states to describe quantized discrete spatial geometries~\cite{oriti2015group, Chirco:2017vhs, Colafranceschi:2021acz}.

Consider a given graph $\Gamma$. To each edge $e$ of $\Gamma$ we associate a half-integer spin variable $j_e$ labeling a $(2j_e+1)$-dimensional $SU(2)$ irreducible representation space $\hilbert{j_e}$. At the same time, each $N$-valent node $n$ carries an intertwiner state $\ket{I_n}$ in the $SU(2)$-invariant Hilbert space $\mathcal{H}_I=\text{Inv}_{\text{SU}(2)}\left[\bigotimes_{i=1}^N\hilbert{j_i}\right]$, which is the degeneracy space associated to the recoupling of the $N$ spins meeting at the node into a singlet (gauge invariant) representation. A \emph{spin-network} basis state is the triple $\ket{\Gamma; \{j_e\}; \{I_n\}}$, defined by the direct sum over $j_e$ of the tensor product of the gauge invariant states $\ket{I_n}$ at all nodes: 
\begin{equation}
    \ket{\Gamma; \{j_e\}; \{I_n\}}:=\bigoplus_{\{j_e\}}\bigotimes_n \ket{I_n}\,.
\end{equation}
Spin network states can be enriched with a geometric interpretation: each $N$-valent node is dual to a $(N-1)$-simplex, represented at the quantum level by the intertwiner state. 
Accordingly, a 4-valent intertwiner state $\ket{I}$ describes the quantized geometry of a 3-simplex, namely a quantum tetrahedron \cite{BARBIERI1998, PhysRevD.83.044035}.   
In the context of LQG, the presence of clearly defined geometric operators, such as the area operator and the volume operator~\cite{PhysRevD.52.5743, Bianchi_2011}, further strengthens the geometric interpretation; in particular, the volume operator acting on a $4-$valent node is given by 
\begin{equation} 
    \hat V=\left(\frac{\sqrt{2}}{3}\right)^2(8\pi \gamma)^{3}\left(-i\left[\vec J_1\cdot\vec J_2,\vec J_1\cdot\vec J_3\right]\right)\,,
\end{equation}
and has a diagonal representation on the intertwiner basis~\cite{Rovelli_1995}.
Note that this operator is Hermitian but not positive, since it also keeps information of the space orientation \cite{baez1999quantum}. The two possible signs split the degeneracy between the eigenvalues.

For the sake of our work, we focus on the case of a single 4-valent intertwiner with all spins fixed to the same value $j$. The Hilbert space of the tensor product of the four spins $j$ {recoupling into the total spin $J$ can be written} as
\begin{equation}
 \hilbert{j}^{\otimes 4}=\bigoplus_{J=0}^{4j}D^{J}_{j} \hilbert{J}\, , 
\end{equation}
{where the multiplicity spaces (or degeneracy spaces) $D^J_{j}$ consist in the spaces of $SU(2)$-invariant intetrtwiner states in the tensor product of the total spin Hilbert space $\hilbert{J}$ with the individual spins $\hilbert{j}^{\otimes 4}$.}

A state $\ket{I}\in\text{Inv}_{\text{SU}(2)}[\hilbert{j}^{\otimes4}]$ can be written as the recoupling of four spins $\ket{\vec{m}}=\bigotimes_{i=1}^{4}\ket{m_i}\in \hilbert{j}^{\otimes 4}$ into the singlet ($J=0$):
\begin{equation} \label{intertwiner}
\ket{I} =N \sum_{K=0}^{2j}\sum_{M=-K}^K  \sum_{\{\vec{m}\}} C^{K,M}_{jm_1jm_2}C^{K,-M}_{jm_3jm_4}\ket{\vec{m}}\,,
\end{equation}
where $N$ is a normalization factor and $C_{jm_ijm_k}^{KM}$ are the Clebsch-Gordan coefficients involved in the intermediate recoupling of the two pairs of spins $j$ into two states with spin $K$; to ensure the final recoupling into a singlet state, the magnetic indices $M$ have opposite signs. 
We use the shorthand notation $\sum_{\{\vec{m}\}}$ to indicate the sum running over the $(2j+1)^4$ basis element in $\mathcal{H}_{j}^{\otimes 4}$.
In the following, we set all spins to the value $j=1/2$ (see \cref{appqudit} for a short discussion on the case $j\not=1/2$). In this case, the spin representation space reduces to $\hilbert{\frac{1}{2}}=\text{span}\left\{\ket{\uparrow}\equiv\ket{0},\ket{\downarrow}\equiv\ket{1}\right\}\simeq\mathbb{C}^2$, therefore, each spin, which is dual to one of the faces of the tetrahedron, is described by a qubit. The Hilbert space of the tensor product of the four $1/2$-spins decomposes as
\begin{equation}
 \hilbert{\frac{1}{2}}^{\otimes 4}=\bigoplus_{J}D^{J}_{\frac{1}{2}} \hilbert{J}=2\hilbert{0}\oplus 3\hilbert{1}\oplus \hilbert{2}\,.
\end{equation}
The $SU(2)$-invariant subspace $\hilbert{0}$ comes with degeneracy $D^{0}_{1/2}=2$, hence any gauge invariant state $\ket{I}$ is a one qubit state in the 2-dimensional intertwiner Hilbert space 
\begin{equation}
\hi_I:=\text{Inv}_{\text{SU}(2) }[\hilbert{1/2}^{\otimes4}]=\hilbert{0}\oplus\hilbert{0}\simeq\mathbb{C}^2\,,    
\end{equation}
where we can choose a suitable basis $\{\ket{0_s},\,\ket{1_s}\}$. We can represent a 4-valent intertwiner state $\ket{I}\in\hi_I$ both in the computational basis $\{\ket{0}, \ket{1}\}^{\otimes4}$ of the 4-qubits space $\hilbert{1/2}^{\otimes4}$ and in the logical basis $\{\ket{0_s},\ket{1_s}\}\in\hilbert{0}\oplus\hilbert{0}$. In the following we will refer to a state $\ket{I}$ written in the logical basis as a \emph{logical intertwiner qubit} (LIQ) state \cite{mielczarek2018}. 

In terms of the computational basis, the expression of the elements of logical basis can be found using~\eqref{intertwiner}. Explicitly, one finds \cite{Feller_2016,Czelusta_2021}:
\begin{align}
    \nonumber     \ket{0_s}&=\half\left(\ket{0101}+\ket{1010}-\ket{0110}-\ket{1001}\right)\,, \\ 
    \nonumber    \ket{1_s}&=\frac{1}{\sqrt{3}}\big[\ket{0011}+\ket{1100}-\half\big(\ket{0101}\\ & +\ket{1010} +\ket{0110}+\ket{1001}\big)\big]\, \label{intbasis}.
    \end{align}
A generic LIQ state is given by the following Bloch representation
\begin{equation}
    \ket{I(\theta,\phi)}=\cos{\frac{\theta}{2}}\ket{0_s}+\sin{\frac{\theta}{2}}e^{i\phi}\ket{1_s} \, ,
    \label{genericIntertwiner}
\end{equation}
where $\theta\in[0,\pi]$ and $\phi\in[0,2\pi)$ are angles on the Bloch sphere. 
Finally, it can be shown \cite{Czelusta_2021} that the eigenstates of the volume operator take the following form:
\begin{align}
    \ket{V_+}&=\frac{1}{\sqrt{2}}(\ket{0_s}-i\ket{1_s})\,, \\
    \ket{V_-}&=\frac{1}{\sqrt{2}}(\ket{0_s}+i\ket{1_s})\,,
\end{align}
that means $\hat V\ket{V_\pm}=\pm V_0 \ket{V_{\pm}}$. These states represent remarkable points on the Bloch sphere, as they are placed at the equator and their angular coordinates are $(\theta=\frac{\pi}{2},\phi=\frac{\pi}{2})$ and $(\theta=\frac{\pi}{2},\phi=\frac{3\pi}{2})$.

\section{Stabilizer entropy of a quantum tetrahedron} \label{sec:SEqt}
\begin{figure}[ht]
\centering
    \includegraphics[width=0.5 \textwidth]{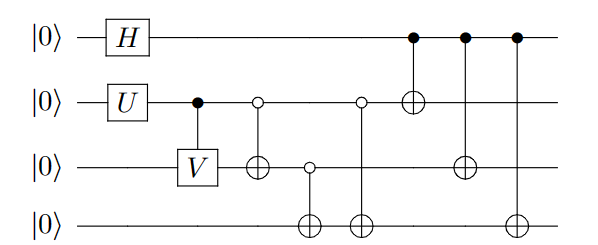}
    \caption{Circuit realization of the intertwiner state $\ket{I(\theta,\phi)}$. $H$ is the Hadamard gate, $U$ and $V$ are specific gates depending on $\theta$ and $\phi$, and the others are CNOT and anti-CNOT gates.}
    \label{fig:intertwinercircuit}
\end{figure}

We shall investigate whether the basis states of the gauge invariant subspace $\hi_I\subset\hilbert{1/2}^{\otimes4}$ possess SE. In this sense, we investigate whether the gates of the quantum circuit associated with the construction of LIQ states $\ket{I}$ from the computational basis belong to the Clifford group.
Without loss of generality, we can start from the reference state $\ket{0}^{\otimes4}$, and then look for unitary transformations such that $\ket{0_s}=\operatorname{U}_{0_s}\ket{0}^{\otimes 4}$ and $\ket{1_s}=\operatorname{U}_{1_s}\ket{0}^{\otimes 4}$.
Using the relations in \cref{intbasis}, we can express the unitary operators $\operatorname{U}_{0_s}$ and $\operatorname{U}_{1_s}$ in terms of a set of unitary gates acting on a stabilizer reference state in the 4-qubit Hilbert space. Thereby, it is possible to realize the generic intertwiner state $\ket{{I(\theta,\phi)}}$ via the quantum circuit given in \cref{fig:intertwinercircuit} \cite{Czelusta_2021} (see also \cite{Girelli_2005} and \cite{MARZUOLI200279, marzuoli} for different descriptions of quantum spin network circuits). The only non-Clifford gates in the circuit are the two special unitary operators, $U$ and $V$, which depend on the parameters $\theta$ and $\phi$ of the output state $\ket{{I(\theta,\phi)}}$.

\begin{figure}[ht]
    \centering
    \includegraphics[width=0.39\textwidth]{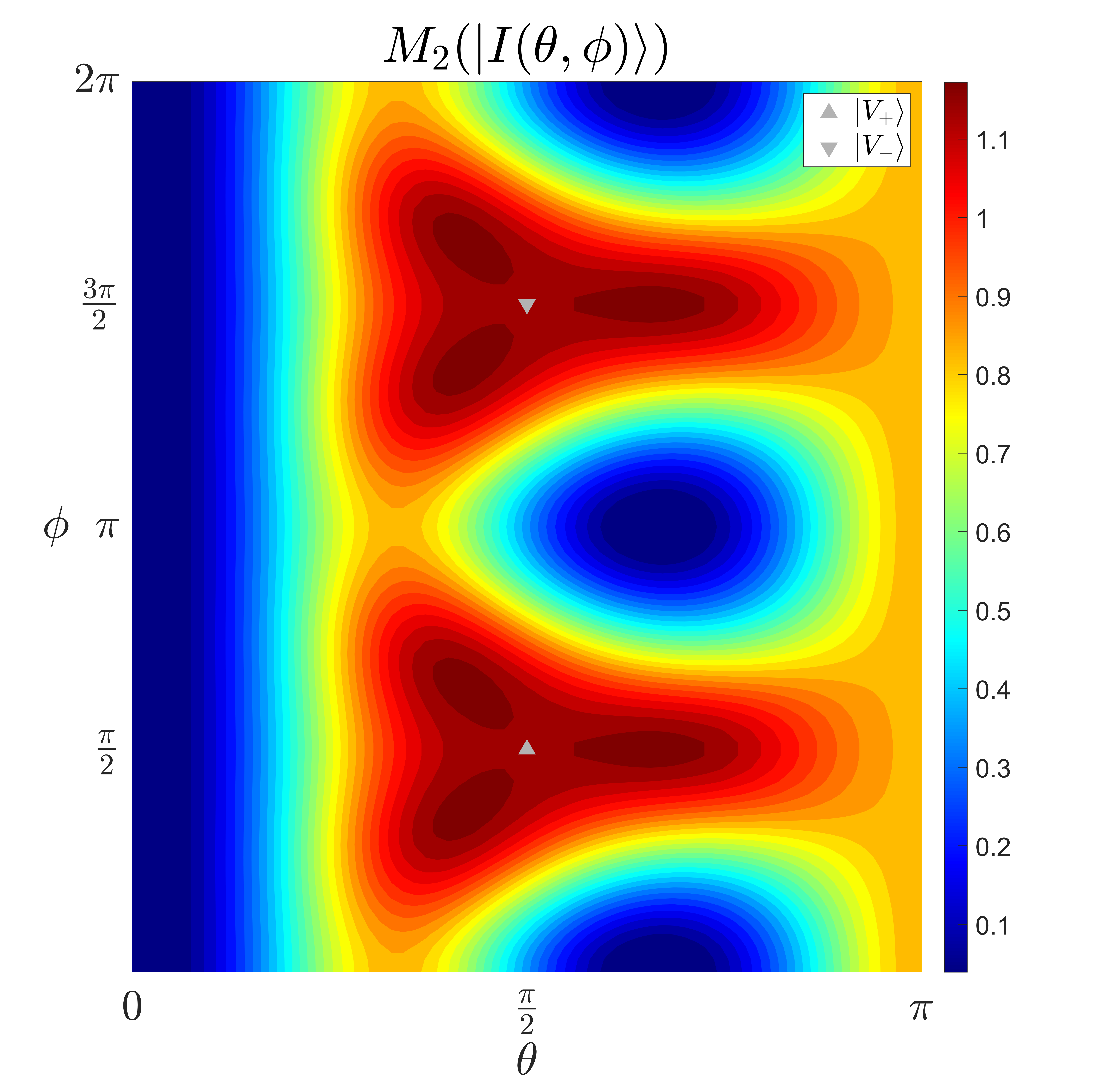}
    \caption{Contour plot of the $2-$Stabilizer Rényi entropy $M_2(\ket{I(\theta,\phi)})$ as a function of the Bloch sphere angles $\theta\in[0,\pi]$ and $\phi\in[0,2\pi)$. The two volume eigenstates, $\ket{V_+}$ and $\ket{V_-}$, are highlighted in the plot.}
    \label{plotmagic}
\end{figure}

Such operators can be written as $2 \times 2$ matrices \cite{Czelusta_2021}:
\begin{equation}
U=
    \begin{pmatrix}
        c_0 & \sqrt{\absval{c_+}^2+\absval{c_-}^2}\\
        -\sqrt{\absval{c_+}^2+\absval{c_-}^2} & c_0^* 
    \end{pmatrix} \,,
\end{equation}
\begin{equation}
V=\frac{1}{\sqrt{\absval{c_+}^2+\absval{c_-}^2}}
    \begin{pmatrix}
        -c_+ & c_-^*\\
        -c_- & -c_+^*
    \end{pmatrix},
\end{equation}
where the coefficients $c_0$ and $c_{\pm}$ are given by the following functions of $\theta$ and $\phi$:
\begin{align}\label{UVcoeff}
    &c_0=\sqrt{\frac{2}{3}}e^{i\phi}\sin{\frac{\theta}{2}}\,,\\&c_\pm=\frac{1}{\sqrt{2}}\left[-\frac{1}{\sqrt{3}}e^{i\phi}\sin{\frac{\theta}{2}}\pm\cos{\frac{\theta}{2}}\right]\,.
\end{align}
Our first remark is that the SE of an intertwiner state $\ket{I(\theta,\phi)}$ seen from the full space is entirely determined by the gates $U$ and $V$, which in general are not Clifford gates. Indeed, the calculation of the SE for the LIQ basis states (see \cref{App1} for full details) gives
\begin{align}\label{m2intbasis}
    &M_2(\ket{0_s})=0\,,\\  \nonumber
    &M_2(\ket{1_s})=0,847997\,. 
\end{align}
Hence, a generic superposition of these two basis states will not be in general a stabilizer state. We can plot the SE of all the intertwiner states in the basis of the Hilbert space $\hilbert{1/2}^{\otimes4}$ as a function of the Bloch sphere representation of the gauge invariant space $\hi_I$ (\cref{plotmagic}).
In particular, we remark the behavior of the volume eigenstates $\ket{V_{\pm}}$: from the SE standpoint, they are located at the center of the regions of maximal non-stabilizerness and they belong to the same Clifford orbit, having exactly the same value of SE:
\begin{equation} \label{m2intvolume}
M_2(\ket{V_{+/-}})= 1,16993\,.
\end{equation}

We recall that the SE of these states is calculated with respect to the full Pauli operator basis of $\hi_{\frac12}^{\ot 4}$, namely the Pauli group of $4$-qubits $\pauli{4}$. This distinction is necessary since, as we have seen in Eq. \eqref{m2intbasis}, the basis states themselves are not stabilizer states in $\hi_{\frac12}^{\ot 4}$, whereas they are stabilizers in $\hi_I$. 
Finally, notice that these states are also non separable in the tensor product structure associated to $\hi_{\frac12}^{\ot 4}$, as a demonstration that {\em both} entanglement and SE are necessary for the gauge structure\cite{hamma2005entanglementbilocal, Ghosh_2015, donnelly2008entanglementloop, schliemann2005entsu2}. 

\section{Average SE of \pdfmath{SU(2)}-gauge invariant subspace}\label{sec:gap}
{The reason for having different values of SE in the two bases states, as shown in Eq. \eqref{intbasis}, is rooted in the gauge structure of the intertwiner state. We shall explore and generalize this result further.} The $SU(2)-$gauge invariant intertwiner space $\hi_I$ is a subspace of the Hilbert space of four qubits $\hi_\half ^{\otimes 4}$. However, the SE in a particular basis state has hardly any physical meaning as any superposition of states in $\hi_I$ is allowed and SE is not constant in any given subspace. To associate SE in a meaningful way to a subspace, one is also confronted with the choice of the Pauli basis with respect to the SE must be computed. To be concrete, if one has a state $\ket{\psi}\in\hi_I$, do we want to know the SE of this state as a state expressed in the computational basis of $\mathcal H^{\otimes 4}$ or in the computational basis of $\hi_I$ given by $\{\ket{0_s},\ket{1_s}\}$? 

In the construction of a quantum gauge theory, one starts with an ambient Hilbert space $\mathcal{H}_{\operatorname{tot}}$. The gauge constraints are expressed as local projectors $\Pi^{(s)}$ and the gauge invariant subspace is the global projection over all the local gauge constraints,
\ba
\hi_G =\Pi_G \hi_{\operatorname{tot}}\,,
\ea
with $\Pi_G:=\prod_s \Pi^{(s)}$.

We associate to a subspace $\hi_G$ its average SE with respect to the Heisenberg-Weyl basis $\{\mathcal{D}^{(i)}\}\subset \mathcal{L}(\hilbert{i})$. In this notation, $i=0$ refers to the Heisenberg-Weyl basis in $\hi_{\operatorname{tot}}$, while $i=G$ refers to the gauge invariant Hilbert space $\hi_G$. With this notation, the linear SE is 
$M^{(i)}_{\lin}(\psi):=1-d_i\tr [Q^{(i)} \psi^{\otimes 4}]\,,i\in\{0,G\}$ and the operator 
\begin{equation}
    Q^{(i)}:= d_i ^{-2}\sum_{(p,q)}D_{(p,q)}^{(i)\ot 4}
\end{equation}
is defined accordingly with the corresponding Heisenberg-Weyl (see \cref{appqudit} for details) basis. The average SE in the subspace $\hi_G$ is then defined as
\ba\label{ave-ns}
M^i &:=& \exv_{U_G} M_{\lin} ^{(i)}(\psi_{U_G}) \\
&=&1-d_i \tr [Q^{(i)} \exv_U \psi_{U_G} ^{\otimes 4}] \,,
\ea
with $\exv_{U_G}$ denoting the unitary group average with respect to the Haar measure over $\mathcal H_G$, 
and $\psi_{U_G}:=U_G\psi U_G^\dag$. In order to perform the Haar average over the subspace $\hi_{G}$ we need the following:
\begin{lemma}\label{avgsublemma}
    Given any Hilbert space $\hi=\hi_R\oplus\hi_R ^\perp$, with $\operatorname{dim}(\hi)=d$ and $\operatorname{dim}(\hi_R)=d_R$, the unitary Haar average of $k$ copies of the state over the subspace $\hi_R$ is given by
    \begin{equation}\label{lemma}
        \exv_{U_R}\psi_{U_R} ^{\ot k}=c_R(d,d_R)\Pi_R^{\ot k}\exv_{U}\psi_{U} ^{\ot k}\,,
    \end{equation}
    with $U_R\in \mathcal{U}(\hi_R)$, $U\in\mathcal{U}(\hi)$, $\Pi_R$ the projector onto $\hi_R$, $\psi\in\hi_R$ and $c_R(d,d_R)=\binom{d+k-1}{k}\binom{d_R+k-1}{k}^{-1}$.
\end{lemma}
{A proof of this lemma is given in \cref{avgsubapp}}. Based on the above lemma, we obtain the general result
\ba\label{lem}
M^i = 1-d_i c_i(d,d_i)\tr [Q^{(i)} \Pi_G^{\ot 4}\exv_{U}\psi_{U} ^{\ot 4}]\,,
\ea
where now $\exv_{U}$ denotes the Haar average over the full  unitary group on $\hi_{\operatorname{tot}}$. 

Eq.\eqref{lem} shows explicitly how the gauge structure enters the SE through the projector $\Pi_G$. When $i=G$, this projector becomes the identity map and the average SE has no recollection of the gauge structure. In order to quantify the amount of SE due to this structure, we define the \textit{SE-gap} as
\ba
    \Delta M(\hi_G)&:=& M^0-M^G\\ 
    \nonumber
    &=&\exv_{U_G} M_{\lin} ^{(0)}(\psi_{U_G})-\exv_{U_G} M_{\lin} ^{(G)}(\psi_{U_G})\,.
\ea

For a general $SU(2)-$gauge structure, namely the spin $j$ intertwiner states, the projector $\Pi_I$
reads (see \cref{appqudit})
\ba
\hspace{-0.5 cm}\Pi_I&=&\ketbra{0_s}+\ketbra{1_s}+\dots+\ketbra{2j_s}
\ea

{Let us now specialize} these formulae to the case of $j=1/2$, that is, the quantum tetrahedron.
Using Lemma \ref{avgsublemma}, with $k=4$, $\hi=\hi_\half ^{\ot 4}$, $\hi_R=\hi_I$ and $\Pi_I=\ketbra{0_s}+\ketbra{1_s}$ being the projector onto the intertwiner space,  (see \cref{mgap} for details) we find
\ba
    M^0 =17/45.
\ea

Now, notice that the logical states $\ket{0_s}, \ket{1_s}$ in the Pauli basis in which $Z_I = \ket{0_s}\bra{0_s}- \ket{1_s}\bra{1_s}$ are obviously stabilizer states with zero stabilizer entropy. Nevertheless even according to this Pauli basis there will be a non-zero value for the average stabilizer entropy. This space is just a generic qubit $\mathbb C^2$ from the point of view of the stabilizer entropy. The average SE in this space is thus just the average stabilizer entropy of a qubit, namely $M^I:=\exv_{U_I} M_{\lin} ^{(I)}(\psi_{U_I})=1/5$, as calculated in \cite{leone2022StabilizerRenyiEntropy} (we also explicitly show this calculation in \cref{mgap}).

Putting the pieces together, we are able to calculate the average SE gap, which reads 
\begin{equation}
    \Delta M(\hi_I)=8/45\,.
\end{equation}
{A value of $\Delta M(\hi_I)$ greater than zero tells us that projecting a generic $4$-qubit state onto this gauge invariant subspace has a cost in terms of non-stabilizer resources.} 

{In particular, this means that the gauge structure bears a cost in terms of \emph{simulability}, which is very important as one scales the system to many nodes.}

\section{Simulations of quantum gravity states}\label{sec:simulations}

{Very recently, quantum gravity states have been physically implemented on quantum computers \cite{Czelusta_2021,vanderMeer:2022jec}, and in particular, quantum tetrahedra states.} The very first layer of difficulty that must be faced in a laboratory when attempting to conduct a quantum experiment (including one regarding simulations of quantum gravity states) is the preparation of an initial state that is faithful to the theoretical one $\ket{\psi}$. 
Typically, a large initial sample must be prepared, resulting in a mixed output $\tilde{\psi}$ from the processor.
Ensuring the correct functioning of quantum devices in terms of the accuracy of the output requires a certification protocol \cite{Eisert_2020}; one of the possible measures of the quality of the realization $\tilde{\psi}$ of a state $\ket{\psi}$ is the fidelity $\mathcal{F}(\ket{\psi},\tilde{\psi})=\tr (\psi\tilde{\psi})$ that measures the precision of preparation. It is known that SE can provide useful indications in an experimental setting.

In this last section, we argue that the numerical results found in \eqref{m2intbasis} and \eqref{m2intvolume} have a direct use in the recent results on quantum gravity states  simulations. Indeed, we can use the SE of these states to estimate the maximum fidelity achievable with a given number of preparations or, conversely, the minimum number of preparations needed to achieve a desired value of fidelity within a desired error \cite{nonstabfidest}.

In \cite{Czelusta_2021}, the authors present a realization of the intertwiner states $\ket{0_s},\ket{1_s}$ as of Eq. \eqref{intbasis} as well as the volume eigenstates $\ket{V_+},\ket{V_-}$ on a $5$-qubit (Yorktown) and a $15$-qubit (Melbourne) IBM superconducting quantum computer. By performing 10 rounds of 1024 quantum measurements, they obtained fidelity values $\mathcal{F}_{exp}$ of the prepared states with respect to the theoretical ones. As anticipated in Sec. \ref{stabform}, the explicit lower bound on the number $N_{\tilde{\psi}}^{\text{min}}$ of preparations needed to achieve an accuracy $\epsilon$, with a probability of failure $\delta$, is related to the SE of the theoretical state $\ket{\psi}$ as follows:  
\begin{equation}\label{nmin}
    N_{\tilde{\psi}}\geq \frac{2}{\epsilon^2}\ln\left(\frac{2}{\delta}\right)\exp[M_2(\ket{\psi})]\,.
\end{equation}

Inverting this relation, one can get an upper bound on the maximum achievable fidelity with a given number of preparations $N$ with failure probability $\delta$:
\begin{equation}\label{fmax}
    \mathcal{F}_{\text{max}}\leq 1-\sqrt{\frac{2}{N}\ln\left(\frac{2}{\delta}\right)}\exp\left[\frac{M_2(\psi)}{2}\right]\,.
\end{equation}

 \begin{table}[h!]
     \centering
     \begin{tabular}{|c|c|c|c|c|c|c|}
     \hline
         $\ket{I}$ & $\mathcal{F}_{\text{exp}}$ & $\sigma$ & $M_2$ & $(1-\delta)$ & $\mathcal{F}_{\text{max}}$ & $N_{\tilde{\psi}}^{\text{min}}$ \\
         \hline
        $\ket{0_s}$ &  0,906 & 0,005 & 0 & 0,05 & 0,973 & 835  \\
         $\ket{1_s}$ & 0,916 & 0,007 & 0,847997 &  0,05 &  0,959 & 2441 \\
         $\ket{V_+}$ & 0,918 & 0,009 & 1,16993 &0,05  & 0,952 & 3535\\
         $\ket{V_-}$ & 0,917  &0,008  & 1,16993 &0,05  & 0,952 & 3450 \\
         \hline
     \end{tabular}
     \caption{Comparison between the experimentally obtained results of the fidelity in \cite{Czelusta_2021} and the bounds given by Eqs. \eqref{nmin} and \eqref{fmax}. $\mathcal{F}_{\operatorname{max}}$ is calculated using $N=10240$ and the probability of failure $\delta$ is inferred using the Bienaymé-Čebyšëv inequality assuming the statistical error $\sigma$ as the variance of the probability distribution.}
     \label{fid}
 \end{table}

As one can see from the Table \ref{fid}, the experimentally obtained data for the the fidelity is perfectly compatible with the bounds provided by the SE. However, the authors used many more copies than the minimum shown in the table. However, it is important to note that the constraints provided by Eqs. \eqref{nmin} and \eqref{fmax} are purely theoretical, hence independent of the inherent noise sources peculiar to the specific implementation protocols and hardware used in the experimental setting at hand. Nevertheless, by considering the SE of the states one wishes to prepare, the hardware resources and the number of preparations can be managed more efficiently in future experiments of quantum gravity states.

\section{Discussion}
In this paper, we have shown that the gauge invariant structure of quantum geometry states has a cost in terms of non-stabilizer resources. This implies that simulations of quantum geometry states can run more efficiently on a quantum computer and that preparations of future experiments can be more efficient if the non-stabilizer property of the state is taken into account. Moreover, we have seen that eigenstates of the oriented volume have  near-maximal amount of SE: this begs the question of why such states possess greater quantum complexity, {suggesting that a correspondence between entanglement and geometry in quantum gravity may extend at a deeper layer of quantumness}. 
Concretely, the first step to answer this question will be to repeat this analysis for a generic spin-$j$ intertwiner and see if, also in that setting, the volume eigenstates are states with maximum SE. A further intriguing direction to explore is the role of non-stabilizer resources when taking into account the quantum state associated to {an actual spin network, that is a collection of intertwiners describing a quantum simplicial complex}: in that general setting, one should expect additional non-stabilizer resources coming from the graph structure, that is the adjacency matrix, describing the connectivity and the non-trivial additional geometrical degrees of freedom described by the holonomies dressing the links. {In particular, in this sense, we expect that non-stabilizerness can be further used to characterise the transition amplitudes, that is the evolution, of quantum geometry states (see e.g. \cite{vanderMeer:2022jec}).}

{In general terms,} the dependence of the SE gap from the projector onto the intertwiner subspace opens to more wide-reaching questions: how does the gauge structure affect non-stabilizerness in a more general setting? Can we use this formalism to characterize the SE  of other quantum gauge theories? Does abelianity (or lack thereof) of the gauge group play a role in the non-stabilizer resources of the gauge invariant subspace?

\section{acknowledgments}
The authors thank L. Leone and L. Vacchiano for important discussions. AH acknowledges support from the PNRR MUR project PE0000023-NQSTI and PNRR MUR project CN $00000013$-ICSC.

\newpage
\appendix
\onecolumngrid

\section{Calculation of the SE of the intertwiner basis states}\label{App1}
In this section, we show the details of the calculations of the SE of the intertwiner basis state, namely the south and north poles of the Bloch sphere (i.e. $\theta=0$ and $\theta=\pi$). Recalling the circuit realization of the intertwiner states in (\cref{fig:intertwinercircuit}), we focus on the unitary operators $U_{0}$, $V_{0}$, $U_{1}$, $V_{1}$ involved in the preparation of basis states $\ket{0_s}$ and $\ket{1_s}$, which can be calculated by inserting in \cref{UVcoeff} $\phi=0$ and $\theta=0,\pi\,$, respectively:
\begin{align}
   & U_{0}=\left(\begin{array}{cc}
        0 &1  \\
         -1&0 
\end{array}\right)\,\,\, &V_{0}=\frac{1}{\sqrt{2}}\left(\begin{array}{cc}
      -1   &-1  \\
         1& -1
    \end{array}\right) \\
     &U_{1}=\sqrt{\frac{2}{3}}\left(\begin{array}{cc}
        1 &\frac{1}{\sqrt{2}}  \\
         -\frac{1}{\sqrt{2}}&1
\end{array}\right)\,\,\, &V_{1}=\frac{1}{\sqrt{2}}\left(\begin{array}{cc}
      1   &-1  \\
         1& 1
\end{array}\right)\label{1sop}
\end{align}

\begin{figure}[ht]
\centering
    \includegraphics[width=0.25\textwidth]{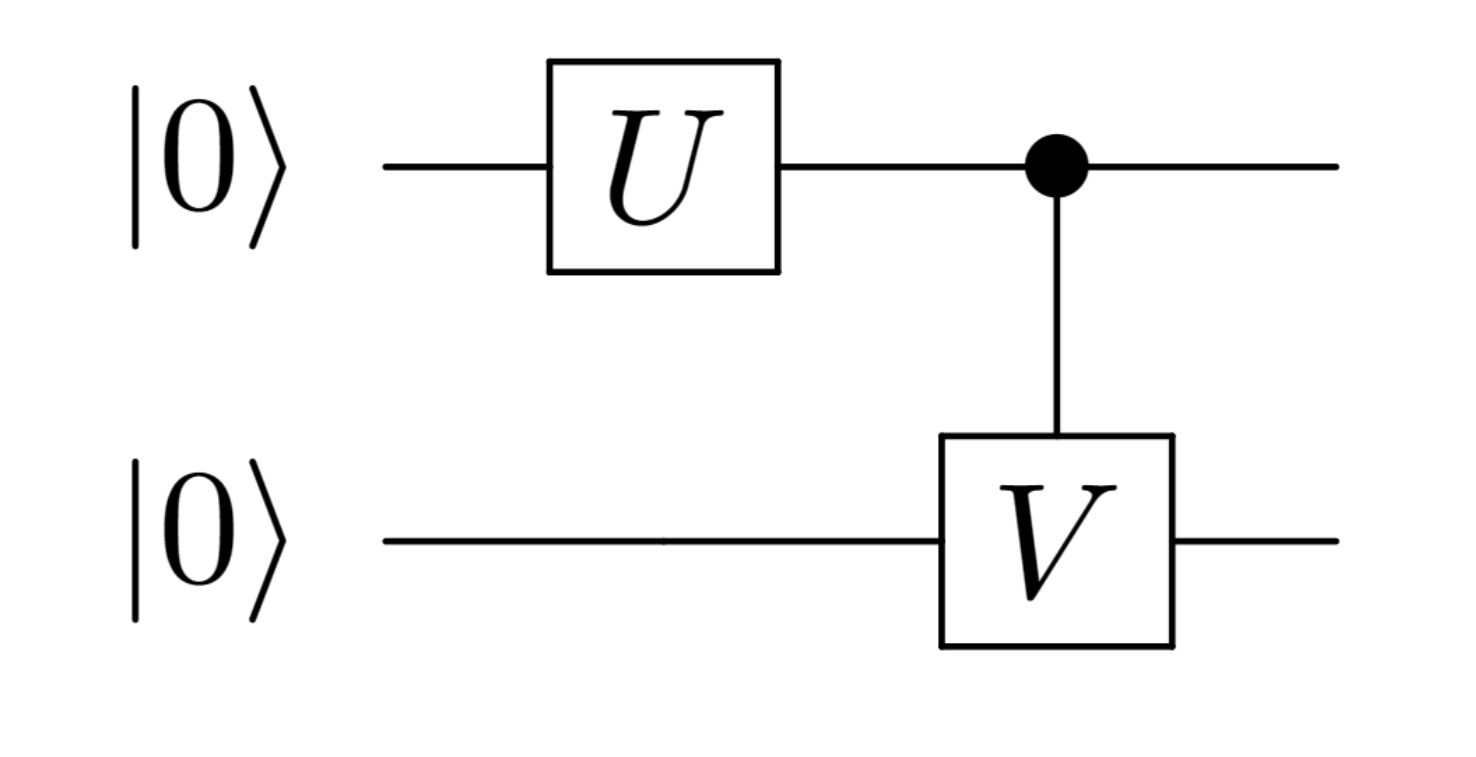}
    \caption{2-qubit circuit with the non-trivial contribution to the magic of an Intertwiner state.}
    \label{2qubitcircuit}
\end{figure}

Notice that, with the exception of the unitaries that we just calculated, all the gates involved in the circuit in (\cref{fig:intertwinercircuit}) are Clifford gates. Hence, to estimate the magic of an intertwiner state, we can focus our analysis on the reduced 2-qubit system given by the action of $CV_0\,(U_0\otimes \Id)$ on $\ket{0}^{\otimes 2}$, where $CV_0$ is the controlled-$V_0$ gate. The reduced circuit is represented in \cref{2qubitcircuit}.

This circuit returns the realization of $\ket{0_s}$ as a 2-qubit state: 
 \begin{equation}
\ket{0_s}_2=\frac{1}{\sqrt{2}}\left(\ket{10}-\ket{11}\right)
 \end{equation} 
In this way, we rule out all trivial contribution to the non-stabilizerness, isolating only the significant ones.

For the particular case of $\ket{0_s}$ we can prove that the magic produced by the operators $U_0$ and $V_0$ is 0; i.e. $\ket{0_s}$ is a stabilizer state. 
Let us write the matrix form of the operators:
\begin{equation}
    U_0\otimes\Id=\left(\begin{array}{cccc}
         0&0&1&0  \\
         0&0&0&1 \\
         -1&0&0&0\\
         0&-1&0&0\end{array} \right)
\end{equation}

\begin{equation}
    CV_0=\left(\begin{array}{cccc}
         1&0&0&0  \\
         0&1&0&0 \\
         0&0&-\frac{1}{\sqrt{2}}&-\frac{1}{\sqrt{2}}\\
    0&0&\frac{1}{\sqrt{2}}&-\frac{1}{\sqrt{2}}\end{array} \right)
\end{equation}

We consider the state $\psi_{0_s}=\ket{0_s}_2\bra{0_s}_2$, which in the Pauli basis reads
 \begin{align}
     \psi_{0_s}&=\frac{1}{d}\sum_{P\in\mathbbm{P}_2}\operatorname{Tr}(\psi_{0_s}P)P \nonumber \\
     &=\frac{1}{4}\left(\Id \otimes \Id-\Id\otimes\X-\Z\otimes\Id+\Z\otimes\X\right) \label{0s}
 \end{align}
The magic of the basis state is
\begin{equation}
    M_2(\psi_{0_s})=-\log d^{-1}\sum_{P\in\mathbb{P}_2}\tr^4(\psi_{0_s}P).
    \label{Magic0s}
\end{equation}
Since the trace of the product of Pauli matrices is equal to $d$ only if the product returns $\Id \otimes \Id$ and $0$ otherwise, among the terms of the sum in \eqref{Magic0s} there are only four non vanishing contributions, which are the ones with $P$ equal to one of the terms in the Pauli decomposition of the state. 
Equation \eqref{Magic0s} returns
\begin{equation}
    M_2(\psi_{0_s})=-\log\left(d^{-1}\left(\frac{d}{4}+\frac{d}{4}+\frac{d}{4}+\frac{d}{4}\right)\right)=0
\end{equation}
Namely, the intertwiner state $\ket{0_s}$ is a stabilizer state.

We now repeat the same procedure for the state $\ket{1_s}$. First, we realize the operators as matrix:
\begin{equation}
    U_1\otimes\Id=\left(\begin{array}{cccc}
         \sqrt{\frac{2}{3}}&0&\frac{1}{\sqrt{3}}&0  \\
         0& \sqrt{\frac{2}{3}}&0& \frac{1}{\sqrt{3}} \\
         - \frac{1}{\sqrt{3}}&0&\sqrt{\frac{2}{3}}&0\\
         0&- \frac{1}{\sqrt{3}}&0& \sqrt{\frac{2}{3}}\end{array} \right)
\end{equation}

\begin{equation}
    CV_1=\left(\begin{array}{cccc}
         1&0&0&0  \\
         0&1&0&0 \\
         0&0&\frac{1}{\sqrt{2}}&-\frac{1}{\sqrt{2}}\\
    0&0&\frac{1}{\sqrt{2}}&\frac{1}{\sqrt{2}}\end{array} \right)
\end{equation}
The action of this operators on $\ket{0}^{\otimes 2}$ returns
\begin{equation}
    \ket{1_s}_2=\sqrt{\frac{2}{3}}\ket{00}-\frac{1}{\sqrt{6}}\ket{10}-\frac{1}{\sqrt{6}}\ket{11}
\end{equation}
We write the state $\psi_{1_s}$ as
\begin{align}
    \psi_{1_s}&=\frac{1}{4}\Big(\Id\otimes\Id+\frac{1}{3}\Id\otimes\X+\frac{2}{3}\Id\otimes\Z-\frac{2}{3}\X\otimes\Id \nonumber \\
    &-\frac{2}{3}\X\otimes\X-\frac{2}{3}\X\otimes\Z+\frac{2}{3}\Y\otimes\Y+\frac{1}{3}\Z\otimes\Id \nonumber \\
    &-\frac{1}{3}\Z\otimes\X+\frac{2}{3}\Z\otimes\Z\Big) \label{1s}
\end{align}
There are ten non-vanishing contributions to the magic of this state, each of which is equal to the fourth power of one of the coefficients of \eqref{1s}. 
Direct calculation returns 
\begin{equation}
    M_2(\psi_{1_s})=0,847997
\end{equation}

\section{Average SE gap}\label{mgap}
In this section, we show the details of the calculations of the average SE gap shown in \cref{sec:gap},
\begin{equation}
    \Delta M(\hi_I):=\exv_U M_{\lin} ^{(0)} (\psi_U)-\exv_U M_{\lin} ^{(I)} (\psi_U)
\end{equation}
with $\exv_U$ denoting the unitary group average with respect to the Haar measure, $M_{\lin}^{(i)}(\psi) :=1-d_i\tr Q^{(i)} \psi^{\otimes 4}\,,i=\{0,I\}$ being the linear SE, $d_i$ being the dimensions of the Hilbert spaces involved (recall that $d_I$ is the dimension of the gauge invariant intertwiner space, hence $d_I=2$, whereas $d_0$ is the dimension of the ambient $4$-qubit space, thus $d_0=16$), and
\begin{equation}
    \begin{split}
        Q^{(I)}&=\frac{1}{d_I^2}\sum_{P\in \pauli{1}} P^{\otimes 4}\\
        Q^{(0)}&=\frac{1}{d_0^2}\sum_{P\in \pauli{4}} P^{\otimes 4}
    \end{split}
\end{equation}
with $\pauli{n}$ being the Pauli group on $n$ qubits modulo the phases. 

\subsection{Haar averages and calculation of $M^I$}
We start by calculating $\exv_U M_{\lin} ^{(I)} (\psi_U)$ since it is the simplest one: plugging the definition of $M_{\lin}$ in the Haar average reads

\begin{equation}
    \exv_U M_{\lin} ^{(I)} (\psi_U)=\exv_U 1-d_I \tr Q^{(I)}\psi_U ^{\otimes 4}=1-d_I \tr Q^{(I)}\exv_U U^{\otimes 4}\psi^{\ot 4} U^{\dag \otimes 4}
\end{equation} 
since the Haar average is linear. Our focus, then, is on evaluating 
\begin{equation}\label{spg}
   \operatorname{SP}^{(I)}:=\exv_U \tr Q^{(I)}\psi_U ^{\otimes 4}
\end{equation}

In general, carrying out the Haar average of operators of the form $U^{\otimes 4} X U^{\dag \otimes 4}\,, X\in L(\hi^{\otimes k})$ requires the knowledge of the commutant of the $k$-tensored representation of the unitary group, according to Schur's Lemma. By the Schur-Weyl duality, the basis of the full commutant of $U^{\ot k}$ is constituted by the permutation operators acting over the $k$ copies of the Hilbert space of interest. In particular, the average of the $k$ copies of a state is carried out in detail in \cite{roberts2017ChaosComplexityDesign}, and reads

\begin{equation}
    \exv_U U^{\otimes k}\psi^{\otimes k} U^{\dag \ot k}=\binom{d+k-1}{k}^{-1}\Pi_{\sym}^{(k)}\,,
\end{equation}
with
\begin{equation}
    \Pi_{\sym} ^{(k)}:=\frac{1}{k!}\sum_{\pi \in S_k} \sswap_\pi
\end{equation}
being the projector onto the subspace of $\hi^{\ot k}$ which is symmetric under permutations of $k$ objects. 
The result of this average is to be expected by the fact that operators belonging $L(\hi^{\ot k})$ of the form $\psi^{\ot k}$ are actually symmetric under permutation operators, so the weight associated to each of this operators must be the same and is hence only determined by the normalization.

The permutation operators relative to $\pi\in S_k$ can be written in the computational basis of $\hi^{\ot k}$, namely $\{\ket{i_1 \dots i_k}\}_{i_h=1} ^{d}$ in this way:
\begin{equation}\label{perm}
    \sswap_{\pi}=\sum_{i_1,\dots,i_k}\ket{\pi(i_1)\dots\pi(i_k)}\hspace{-0.1 cm}\bra{i_1\dots i_k}\,.
\end{equation}
Notice that the permutation operators are invariant under $k$-copies of unitary operators $U^{\ot k}$: this means that such operators will have the same expression in any basis of $\hi^{\ot k}$.
Now, starting 
In our case, we are interested in the Haar average of four copies of the state, namely
\begin{equation}\label{avgpsi4}
\begin{split}
    \exv_{U} \psi_{U} ^{\ot 4}&=\binom{d+3}{4}^{-1}\frac{1}{4!}\sum_{\pi \in S_4} \sswap_\pi \\
    &=[(d+3)(d+2)(d+1)d]^{-1}\sum_{\pi\in S_4}\sswap_\pi    \,.
\end{split}
\end{equation}
Substituting this expression in Eq. \eqref{spg} we get
\begin{equation}
    \begin{split}
        \exv_U\tr Q^{(I)}\psi_U ^{\otimes 4}&=[(d_I+3)(d_I+2)(d_I+1)d_I]^{-1}\tr Q^{(I)}\Pi_\sym ^{(I)}\\
        &=[(d_I+3)(d_I+2)(d_I+1)d_I]^{-1}\frac{1}{d_I^2}\sum_{\pi\in S_4}\sum_{P\in \pauli{2}}\tr P^{\ot 4}\sswap_\pi ^I\,.
    \end{split}
\end{equation}
In order to tackle this calculation, one calculates the sums of the Pauli operators permutation by permutation: by means of example, we show the calculation for one permutation, but the treatment is similar for all of them. Let's take, say, the $3$-cycle $(123)$: the permutation operator associated to this permutation reads
\begin{equation}
    \sswap_{(123)}=\sum_{ijkl}\ket{kijl}\hspace{-0.1 cm}\bra{ijkl}\,,
\end{equation}
hence
\begin{equation}\label{t123}
    \begin{split}
        \tr Q^{(I)}\sswap_{(123)}&=\frac{1}{d_I ^2}\sum_{P\in \pauli{2}}\sum_{ijkl}\tr P^{\otimes 4}\ket{kijl}\hspace{-0.1 cm}\bra{ijkl}\\
        &=\frac{1}{d_I ^2}\sum_{P\in \pauli{2}}\sum_{ijk} \tr(P \ket{k}\hspace{-0.1 cm}\bra{i})\tr(P \ket{i}\hspace{-0.1 cm}\bra{j})\tr(P \ket{j}\hspace{-0.1 cm}\bra{k})\sum_l\tr(P \ket{l}\hspace{-0.1 cm}\bra{l})\\
        &=\frac{1}{d_I ^2}\sum_{P\in \pauli{2}}\sum_{ijk} \tr(P \ket{k}\hspace{-0.1 cm}\bra{i})\tr(P \ket{i}\hspace{-0.1 cm}\bra{j})\tr(P \ket{j}\hspace{-0.1 cm}\bra{k})\tr P\\
        &=\frac{1}{d_I}\sum_{ijk}\tr( \ket{k}\hspace{-0.1 cm}\bra{i})\tr( \ket{i}\hspace{-0.1 cm}\bra{j})\tr( \ket{j}\hspace{-0.1 cm}\bra{k})\\
        &=\frac{1}{d_I}\sum_{ijk}\delta_{ki}\delta_{ij}\delta_{jk}=\frac{1}{d_I}\sum_{jk}\delta_{jk}=1\,,
    \end{split}
\end{equation}
where we used the fact that $\sum_l \ketbra{l}=\Id $ and $\tr P=d_I \delta_{P,\Id}$. Executing similar calculations for the other $23$ elements of $S_4$ and summing all the contributions, one gets
\begin{equation}
    \exv_U\tr Q^{(I)}\psi_U ^{\otimes 4}=\frac{4 d_I ^2+12 d_I+8}{d_I (d_I+1) (d_I+2) (d_I+3)}=\frac25\,,
\end{equation}
hence
\begin{equation}\label{m2i}
    \exv_U M_{\lin}^{(I)}(\psi_U)=1-d_I \exv_U\tr Q^{(I)}\psi_U^{\otimes 4}=\frac15\,. 
\end{equation}


\subsection{Haar Averages on subspaces: proof of Lemma \ref{avgsublemma}}\label{avgsubapp}
Let generic $\hi=\hi_R \oplus \hi_R ^{\perp}$ Hilbert space decomposed in two orthogonal subspaces, with $\operatorname{dim}(\hi)=d$, $\hi_R=\operatorname{span}\{\ket{1},\dots,\ket{d_R}\}$ and $\hi_R ^{\perp}=\operatorname{span}\{\ket{d_R+1},\dots,\ket{d}\}$, and the projector $\Pi_R$ onto $\hi_R$ being
\begin{equation}
    \Pi_R=\sum_{i=1}^{d_R}\ketbra{i}\,.
\end{equation}
Applying the formula shown in Eq. \eqref{avgpsi4}, the average onto the subspace $\hi_R$ reads
\begin{equation}\label{avgR}
        \exv_{U_R} \psi_{U_R} ^{\ot k}=\binom{d_R+k-1}{k}^{-1}\frac{1}{k!}\sum_{\pi \in S_k} \sswap_\pi ^R \,,
\end{equation}
whereas if we calculate the average onto the full Hilbert space $\hi$ we get

\begin{equation}\label{avgfull}
        \exv_{U} \psi_{U} ^{\ot k}=\binom{d+k-1}{k}^{-1}\frac{1}{k!}\sum_{\pi\in S_k}\sswap_\pi\,.
\end{equation}
Now, it suffices to check that the action of $k$ copies of the projector onto the permutation operators representation on the full Hilbert space gives the permutation operators representation on the subspace $\hi_R$. Applying the expression shown in Eq. \eqref{perm} for the permutation operators to this case we get 
\begin{equation}\label{permsub}
    \begin{split}
        \Pi_R^{\ot k} \sswap_\pi&=\sum_{\{i_1,\dots,i_k\}\in \{1,d\}^{\times k}}\Pi_R \ket{\pi(i_1)}\dots\Pi_R \ket{\pi(i_k)}\hspace{-0.1 cm}\bra{i_1\dots i_k}\,.
    \end{split}
\end{equation}
One then notices that 
\begin{equation}
\Pi_R\ket{i}=    \begin{cases}  
    \ket{i}\quad \text{if}\; i\in\{1,d_R\}\\
    0\quad \text{otherwise}
    \end{cases}\,,
\end{equation}
hence Eq. \eqref{permsub} reduces to

\begin{equation}
    \Pi_R^{\ot k} \sswap_\pi=\sum_{\{i_1,\dots,i_k\}\in \{1,d\}^{\times k}}\ket{\pi(i_1\dots i_k)}\hspace{-0.1 cm}\bra{i_1\dots i_k}=\sswap_\pi ^R\,.
\end{equation}
Using this result, Eq. \eqref{avgR} reads

\begin{equation}\label{avgsub}
    \begin{split}
         \exv_{U_R} \psi_{U_R} ^{\ot k}&=\binom{d+k-1}{k}^{-1}\binom{d+k-1}{k}^{-1} \frac{1}{k!}\sum_{\pi\in S_k}\sswap_\pi ^R\\
        &=\binom{d_R+k-1}{k}^{-1} \frac{1}{k!}\Pi_R^{\ot k}\sum_{\pi\in S_k}\sswap_\pi\\
        &=\binom{d+k-1}{k}\binom{d_R+k-1}{k}^{-1}\exv_{U} \psi_{U} ^{\ot k}\\
        &\equiv c_R(d,d_R)\exv_{U} \psi_{U} ^{\ot k}\,,
    \end{split}
\end{equation}
which completes the proof.\qed


\subsection{Calculation of $M^0$}

In this subsection, we use the result of Eq. \eqref{avgsub} to calculate $\exv_U M_{\lin}^{(0)}(\psi_U)$, which reads
\begin{equation}\label{avgm0}
    \exv_U M_{\lin}^{(0)}(\psi_U)=1-d_0\exv_{U_I \in \mathcal{U}(\hi_I)} \tr Q^{(0)} U_I ^{\ot 4} \psi_I ^{\ot 4} U_I ^{\dag\ot 4}\,,
\end{equation}
and in particular, we focus on the evaluation of
\begin{equation}\label{trq0}
    \operatorname{SP}^{(0)}:=\tr Q^{(0)} \exv_{U_I \in \mathcal{U}(\hi_I)}U_I ^{\ot 4} \psi_I ^{\ot 4} U_I ^{\dag \ot 4}\,,
\end{equation}

Using the formula in Eq. \eqref{avgsub}, Eq. \eqref{trq0} reads

\begin{equation}
    \operatorname{SP}^{(0)}=c_I\tr Q^{(0)} \Pi_I^{\ot 4}\exv_{U\in\mathcal{U}(\hi_\half ^{\ot 4})}U^{\ot 4}\psi^{\ot 4}U^{\dag\ot 4} \,,
\end{equation}
with $\Pi_I=\ketbra{0_s}+\ketbra{1_s}$. We can plug the generic formula shown in Eq. \eqref{avgpsi4} and getting

\begin{equation}
    \begin{split}
        \operatorname{SP}^{(0)}&=c_I[d(d+1)(d+2)(d+3)]^{-1}\tr Q^{(0)}\Pi_I^{\ot 4}\sum_{\pi\in S_4}\sswap_\pi\\
        &=[(d_I+3)(d_I+2)(d_I+1)d_I]\tr Q^{(0)}\Pi_I^{\ot 4}\sum_{\pi\in S_4}\sswap_\pi ^I
    \end{split}
\end{equation}

The relevant difference between this calculation and that carried out in Eq. \eqref{m2i} is the representation of the permutation operators $\sswap ^I _\pi$: they now read

\begin{equation}
    \sswap_\pi ^I=\sum_{i_s,j_s,k_s,l_s}\ket{\pi(i_s j_s k_s l_s)}\hspace{-0.1 cm}\bra{i_s j_s k_s l_s}\,.
\end{equation}

with all indices belonging to $\{\ket{0_s},\ket{1_s}\}$ as written in Eq. \eqref{intbasis}. These are indeed operators acting on $\hi_\half ^{\ot 4}$, but they are not the full permutation operators of $\hi_\half ^{\ot 4}$, since the indices of the sums do not run on all the basis elements of $\hi_\half ^{\ot 4 }$, but only on the intertwiner basis elements. Substituting this expression (and the one in Eq. \eqref{avgsub}) in Eq. \eqref{trq0} we get 

\begin{equation}
    \operatorname{SP}^{(0)}=[(d_I+3)(d_I+2)(d_I+1)d_I]^{-1}\frac{1}{d_0^2}\sum_{\pi\in S_4}\sum_{P\in \pauli{4}}\tr P^{\ot 4}\sswap ^I_\pi
\end{equation}

The previous observation renders the calculation of objects like $\tr Q^{(0)}\sswap_\pi ^I$ slightly more difficult, since we cannot exploit the completeness relationship of basis elements like we did in Eq. \eqref{t123}, simply because $\{\ket{0_s},\ket{1_s}\}$ is not a complete basis for $\hi_\half ^{\ot 4}$. However, by pure brute-force methods, we are able to evaluate this object. We proceed permutation by permutation, as before: the trace of $Q^{(0)}$ with a single permutation operator reads
\begin{equation}\label{trq0pi}
    \begin{split}
        \tr Q^{(0)}\sswap^{(s)} _\pi=\frac{1}{d_0 ^2}\sum_P \sum_{i_s j_s k_s l_s}\bra{\pi(i_s)}P\ket{i_s}\hspace{-0.1 cm}\bra{\pi(j_s)}P\ket{j_s}\hspace{-0.1 cm}\bra{\pi(k_s)}P\ket{k_s}\hspace{-0.1 cm}\bra{\pi(l_s)}P\ket{l_s}\,.
    \end{split}
\end{equation}

This sum is constituted by products of matrix elements of $16\times 16$ matrices between two intertwiner basis elements, which are $16$-component vectors in the original $\hi_\half ^{\ot 4 }$ basis. By separately calculating the four possible matrix elements for each and every of the $256$ operators of $\pauli{4}$ (namely, $\{\bra{0_S}P\ket{0_s},\bra{0_S}P\ket{1_s},\bra{1_S}P\ket{0_s},\bra{1_S}P\ket{1_s}\}_{P\in \pauli{4}}$), and combining them according to the permutation $\pi$, and them summing over the Pauli operators $P$, one is able to compute objects of the form \eqref{trq0pi}. Repeating this method for the $24$ permutation operators of $S_4$ and summing the results one gets
\begin{equation}
    \exv_U \tr Q^{(0)}\psi_U ^{\ot 4}=\frac{7}{180}\,,
\end{equation}
hence
\begin{equation}
    \exv_U M^{(0)}_{\lin}(\psi_U)=1-d_0\tr Q^{(0)}\psi_U ^{\ot 4}=\frac{17}{45}\,,
\end{equation}
and finally, we can evaluate the average SE gap, which reads
\begin{equation}
    \Delta M(\hi_I)=\exv_U M^{(0)}_{\lin}(\psi_U)-\exv_U M^{(I)}_{\lin}(\psi_U)=\frac{17}{45}-\frac15=\frac{8}{45}\,.
\end{equation}

\section{SE of 4-valent intertwiner with generic spin}\label{appqudit}
In this section, we introduce a generalized version of the Stabilizer Entropy for qudit systems, following the lines of \cite{Wang2023}. This generalization is needed since the dimension of the intertwiner Hilbert space associated with a quantum tetrahedron with all spins equal to $j\in\frac{\mathbb{N}}{2}$ is $2j+1$, according to Peter-Weyl's theorem. 
Indeed, we can write explicitly the intertwiner state in \eqref{intertwiner} as
\ba
\ket{I}&=&N \sum_{K=0}^{2j}\sum_{M=-K}^K  \sum_{\{\vec{m}\}} C^{K,M}_{jm_1jm_2}C^{K,-M}_{jm_3jm_4}\ket{\vec{m}} \nonumber \\
&=&N\bigg(\sum_{\{\vec{m}\}} C^{0,0}_{jm_1jm_2}C^{00}_{jm_3jm_4}\ket{\vec m}+\sum_{M=-1}^1\sum_{\{\vec{m}\}}C^{1,M}_{jm_1jm_2}C^{1,-M}_{jm_3jm_4}\ket{\vec m} \nonumber \\ 
&+&\ldots +\sum_{M=-2j}^{2j}\sum_{\{\vec{m}\}}C^{2j,M}_{jm_1jm_2}C^{2j,-M}_{jm_3jm_4}\ket{m}\bigg) \nonumber \\
&=& N\big(\ket{0_s}+\ket{1_s}+\ldots +\ket{2j_s}\big)
\ea
where the states $\{\ket{0_s},\ket{1_s},\ldots,\ket{2j_s}\}$ form a set of mutually orthogonal basis elements in the singlet subspace of the Peter-Weyl decomposition of the original Hilbert space.
Hence, the projector on the gauge invariant subspace is
\ba
\Pi_I&=&\ketbra{0_s}+\ketbra{1_s}+\dots+\ketbra{2j_s}\nonumber \\
&=&\sum_{l=0}^{2j}\Pi_l
\ea

For $j\not = 1/2$, this implies that the intertwiner Hilbert space is not a $2$-level system; thus, the usual formulation of stabilizer entropy for qubit systems cannot be applied to such a case. 

In order to establish a definition of Stabilizer entropy for generic qudit systems, we introduce the generalization of Pauli group, namely the \textit{discrete Heisenberg-Weyl group}. Consider an $l$-dimensional Hilbert space $\mathcal{H}\simeq\mathbb{C}^l\,,\,l\in \mathbb{N}$ and the space of linear operators acting on it, namely $\mathcal{L}(\mathcal{H})$: we define the boost and shift operators $X,Z\in \mathcal{L}(\mathcal{H})$ as
\begin{align}
X\ket{j}&=\ket{j\oplus_l 1} \\
Z\ket{j}&=\omega^j\ket{j}\,,
\end{align}
with $\omega=e^{\frac{2\pi i}{l}}$.
The Heisenberg-Weyl operators are defined as
\be
D_{(p,q)}=\tau^{-pq}Z^{p}X^{q}
\ee
with $\tau=e^{\frac{i}{l}\pi}$.
They form a basis for $\mathcal{L}(\mathcal{H})$, given by the $l^2$ operators $\{D_{(0,0)}=Id_{l^2},\, D_{(p,q)}\lvert\,\,\, p,q\in\mathbb{Z}_l\}$, where the orthogonality relation reads
\be 
\tr (D_{(p,q)}D^\dag_{(p',q')})=l\delta_{p,p'}\delta_{q,q'}.
\ee
Finally, the Heisenberg-Weyl group is defined as the group generated by such operators:
\be
\mathcal{D}^{(l)} _1=\left<\{D_{(p,q)}\}\right>
\ee
Notice that for $l=2$ the Heisenberg-Weyl group reduces to the usual Pauli group for qubit systems. 

The Heisenberg-Weyl group acting on a $n-$qudit system is simply given by the tensor product of $n$ copies of the Heisenberg-Weyl group for one qudit. 

\be
\mathcal{D}^{(l)} _n= \mathcal{D}^{(l)\ot n} _1
\ee

Any $n-$qudit state $\psi\in\mathcal{H}^{\ot n}$ can be written in the Heisenberg-Weyl basis as follows
\be
\psi=\frac{1}{d}\sum_{\vec p\in\mathbb{Z}_l^{n}}\sum_{\vec q\in \mathbb{Z}_l^{n}}\tr(\psi D_{(\vec p,\vec q)})D_{(\vec p,\vec q)}\,,
\ee 
with $d:=\operatorname{dim}(\hi^{\ot n})=l^n$.
We can define the normalized expected value of a $n-$qudit pure state over the discrete Heisenberg–Weyl operators as  $\Xi_{D_{(\vec{p},\vec q)}}(\ket{\psi}):=d^{-1}\bra{\psi}D_{(\vec p,\vec q)}\ket{\psi}^2$ and the $\alpha$-Stabilizer R\'enyi Entropy on qudit systems reads
\be
M_{\alpha}(\ket{\psi})=(1-\alpha)^{-1} \log \sum_{\vec p\in\mathbb{Z}_l^{n}}\sum_{\vec q\in \mathbb{Z}_l^{n}} \Xi_{D_{(\vec p,\vec q)}}^{\alpha}(\ket{\psi})-\log d\,.
\ee

In order to compute the SE of a quantum tetrahedron state with $j\not =1/2$, we can refer to \eqref{intertwiner} and write the intertwiner density matrix and its tomography in the Heisenberg-Weyl basis 
\begin{align}
I=\sum_{K,K',M,M'} \sum_{\{\vec{m}\}\{\vec{m}'\}} &C^{*\,K',M'}_{jm'_1jm'_2}C^{*\,K',-M'}_{jm'_3jm'_4} C^{K,M}_{jm_1jm_2}C^{K,-M}_{jm_3jm_4}\ket{\vec{m}}\bra{\vec{m}'}\, , \\
I=\frac{1}{d}\sum_{D_{(\vec p, \vec q)}\in D_4}\sum_{K,K',M,M'} \sum_{\{\vec{m}\}\{\vec{m}'\}}&C^{*\,K',M'}_{jm'_1jm'_2}C^{*\,K',-M'}_{jm'_3jm'_4} C^{K,M}_{jm_1jm_2}C^{K,-M}_{jm_3jm_4} \tr(\ket{\vec{m}}\bra{\vec{m}'} D_{(\vec p,\vec q)})D_{(\vec p,\vec q)}\, ,
\end{align}
where the sums over $K$ and $K'$ run from $0$ to $2j$ and the ones over $M$ and $M'$ run respectively from $-K$ to $K$ and from $-K'$ to $K$; $d=(2j+1)^4$ is the dimension of the Hilbert space over which the trace is performed.

\end{document}